\documentclass[12pt]{article}
\usepackage[cp1251]{inputenc}
\usepackage{amsmath}
\usepackage{amsfonts}
\usepackage{amssymb}
\def\pa{\partial}                       
\def\beq{\begin{eqnarray}}    
\def\eeq{\end{eqnarray}}      

\begin{document}
\date{}
\begin{center}
{\Large\textbf{Cubic vertices of interacting massless spin 4  and
real scalar  fields in unconstrained formulation}}

\vspace{18mm}

{\large
P.M. Lavrov$^{(a, b)} \footnote{E-mail: lavrov@tspu.edu.ru}$
}

\vspace{8mm}

\noindent  ${{}^{(a)}} ${\em
Center of Theoretical Physics, \\
Tomsk State Pedagogical University,\\
Kievskaya St.\ 60, 634061 Tomsk, Russia}

\noindent  ${{}^{(b)}} ${\em
National Research Tomsk State  University,\\
Lenin Av.\ 36, 634050 Tomsk, Russia}

\vspace{20mm}

\begin{abstract}
\noindent
New method (arXiv:2104.11930)
is applied to construct cubic interactions of massless spin 4 and real scalar fields.
In contrast with the case of massless spin 3 fields
(arXiv:2208.05700, 2209.03678, 2210.02842)
the procedure requires to use an unconstrained formulation
(arXiv:0702.161) for the Fronsdal theory.
It is shown that in the unconstrained
formulation  there exists
a four-parameter family of cubic interactions between massless spin 4 and
real scalar fields, which
contains four derivatives in vertices and is invariant with respect
to the original gauge transformation. These vertices contain
cubic interactions between auxiliary  and scalar fields too.
Eliminating all auxiliary fields from the  obtained result
using the equations of motion for the initial action gives a
one-parameter family of cubic vertices for  constrained  massless spin 4
and real scalar fields,
Such cubic vertices are invariant
with respect to constrained gauge transformations of the Fronsdal theory.

\end{abstract}

\end{center}

\vfill

\\

\newpage

\section{Introduction}

The study of interactions among higher spin fields has a long and reach story beginning with
famous papers by Fradkin and Vasiliev \cite{FV1,FV2,Vas1,Vas2,Vas3}. The list of publications
devoted to numerous problems of high spin theory contains hundreds of papers
and there is no way to mention them all here. The simplest interactions involving
higher spin fields are described by cubic vertices including
at least one higher spin field.\footnote{For  recent activities in this sphere see,
for example,
\cite{FKM19,JT19,ZinKh20,BKTM21,BR,BKS22} and references therein.}
Explicit construction of cubic vertices is usually based on the Noether procedure when
an free initial action and gauge transformations
are subjected to deformations maintaining
the gauge invariance of the deformed
action under the deformed gauge transformations in each order
in the deformation parameter. In a certain sense, such the scheme can be considered
as a deformation procedure built into the
Batalin-Vilkovisky (BV) formalism \cite{BV,BV1} and described for the first time
by Barnich and Henneaux in the papers  \cite{BH,H}.
The method \cite{BH,H} operates  infinite number of equations appeared
in expansion of  solutions to the classical master equation of the BV-formalism
with respect to the deformation parameter.
Then the system of these equations is analyzed in each order using cohomological methods.

Recently,
the new method for the deformation procedure within the BV-formalism
has been proposed \cite{BL-1,BL-2,L-2022}.
From a formal point of view, the new approach looks like the summation of the Taylor
expansion in exact and closed form for both the deformed action
and the deformed gauge symmetry.
An informal reason to arrive at the resulting description is related
to the invariance of the classical master equation under
anticanonical transformations, which make it possible
to connect two solutions to each other. In particular, it was allowed to connect
an initial gauge theory with the deformed one with the help of special anticanonical
transformations acting non-trivially in the sector of initial fields \cite{BL-1}.
Applications of the BV-formalism for the deformation procedure require an initial action
to belong the class of unconstrained theories that
means linear independence of initial fields. In the case of the Fronsdal theory
\cite{Fronsdal-1} for free massless integer spin fields it leads to restrictions
on value of spins, $s\leq 3$, when one can use the BV-formalism directly.
The action for massless spin 3 field belongs to the class of first-stage
reducible gauge theories when the BV-formalism works.
Interactions of massless spin 3 field between massive vector and real scalar fields
within the new method have been recently studied in papers \cite{L-ejpc-s3,L-22-2, L-s5}.
It was shown that in all considered models cubic vertices containing one massless
spin 3 field and invariant under original gauge transformations are forbidden
while quartic and quintic vertices can be explicitly constructed. In turn it opened a way
in construction of new consistent models with interactions in Quantum Field Theory.

In the present paper,
the new method \cite{BL-1} is applied to construct cubic interactions of massless
spin 4 and real scalar fields. Such application requires
to use an unconstrained formulation \cite{BGK} for the Fronsdal theory
of massless integer spin fields. It is shown that in the unconstrained
formulation there exists a four-parameter family of cubic interactions
between massless spin 4 and real scalar fields containing four derivatives in vertices
and
being invariant under original gauge transformation. These vertices contain
cubic interactions between auxiliary  and scalar fields too.
Eliminating all auxiliary fields
from the obtained result
using the equations of motion for the initial action
gives an one-parameter family of cubic vertices for constrained
massless spin 4 and real scalar fields. Such cubic vertices are invariant
with respect to constrained gauge transformations of the Fronsdal theory.

The paper is organized as follows. In section 2, gauge invariance of initial action
of massless spin 4 and real scalar fields is discussed.
In section 3, suitable deformations of both initial action and gauge transformations
leading to cubic vertices in the deformed action are described.
In section 4, gauge invariance of local part of the deformed action in unconstrained
formulation is studied. Section 5 is devoted to on-shell limit of local cubic vertices
when all auxiliary fields are extracted with the help of corresponding equations of motion.
In section 6 we summarize the results.

The DeWitt's condensed notations
\cite{DeWitt} are systematically used. The right
functional derivatives are marked by special symbols $"\leftarrow"$.
Arguments of any functional
are enclosed in square brackets $[\;]$, and arguments of any
function are enclosed in parentheses, $(\;)$.

\section{Initial action of model}
\noindent

The action for massless  spin 4 fields within the unconstrained formulation \cite{BGK}
has the form
\beq
\nonumber
&&S_0[\varphi, F, \alpha, \lambda_{(2)}, \lambda_{(4)}]=
\int dx\Big[\varphi_{\mu\nu\lambda\sigma}\Box\varphi^{\mu\nu\lambda\sigma}-
12F_{\mu\nu}\Box F^{\mu\nu}+4\Lambda_{\mu\nu\lambda}\Lambda^{\mu\nu\lambda}+\\
&&\qquad\qquad + \lambda_{(2)}^{\mu\nu}\big(\eta^{\lambda\sigma}\varphi_{\mu\nu\lambda\sigma}-
2F_{\mu\nu}-2\pa_{\mu}\alpha_{\nu}\big)+ \lambda_{(4)}\big(\eta^{\mu\nu}F_{\mu\nu}
-\pa^{\mu}\alpha_{\mu}\big)\Big],
\label{b1}
\eeq
where the notation
\beq
\label{b2}
\Lambda_{\mu\nu\lambda}=\pa^{\sigma}\varphi_{\sigma\mu\nu\lambda}-\pa_{(\mu}F_{\nu\lambda)}
\eeq
is used. In (\ref{b1})
$\varphi^{\mu\nu\lambda\sigma}=\varphi^{\mu\nu\lambda\sigma}(x)$ is completely
symmetric forth rank tensor,
$\Box$ is the D'Alembertian, $\Box=\pa_{\mu}\pa^{\mu}$,  $\eta_{\mu\nu}$ is
the metric tensor of flat Minkowski space of the dimension $d$, and
$F_{\mu\nu}, \alpha_{\mu}, \lambda_{(2)}^{\mu\nu},\lambda_{(4)} $ are auxiliary
 fields of corresponding ranks coinciding with number of indices.

We are going to study a  possibility in construction of interactions
of massless spin 4 fields with a real scalar field $\phi=\phi(x)$.
We assume that the initial action has the form \beq \label{c1}
S_0[A]=S_0[\varphi, F, \alpha, \lambda_{(2)},
\lambda_{(4)}]+S_0[\phi], \eeq where  $S_0[\phi]$ is the action of a
free real scalar field, \beq \label{c1a} S_0[\phi]=\int
dx\frac{1}{2}\big[ \pa_{\mu}\phi\;\pa^{\mu}\phi-m^2\phi^2\big], \eeq
and $A$ denotes collection of all fields of the model,
$A^i=(\varphi^{\mu\nu\lambda\sigma},  F^{\mu\nu}, \alpha^{\mu},
\lambda_{(2)}^{\mu\nu}, \lambda_{(4)}, \phi)$. The action (\ref{c1})
is invariant under the gauge transformations\footnote{The symbol
$(\cdots)$ means the cycle permutation of indexes involved.} \beq
\label{c2}
\delta\varphi^{\mu\nu\lambda\sigma}=\pa^{(\mu}\xi^{\nu\lambda\sigma)},
\quad \delta F^{\mu\nu}=\pa_{\lambda}\xi^{\lambda\mu\nu},\quad
\delta\alpha^{\mu}=\eta_{\lambda\nu}\xi^{\lambda\nu\mu},\quad
\delta\lambda_{(2)}^{\mu\nu}=0, \quad \delta\lambda_{(4)}=0,\quad
\delta \phi=0,
\eeq
where $\xi^{\mu\nu\lambda}$ are arbitrary totally symmetric third rank tensor.
Algebra of gauge transformations is Abelian.

\section{Deformations of initial theory}

Now, we consider possible deformations of initial action using the procedure
which is ruled out
by the generating functions $h^i(A)$ \cite{BL-1}. Here, we restrict ourself by the case of
anticanonical transformations acting effectively in the sector of fields
$\varphi^{\mu\nu\lambda\sigma}$ and $F^{\mu\nu}$ of the initial theory.
It means the following structure of
generating functions
$h^i(A)=(h_{(\varphi)}^{\mu\nu\lambda\sigma}(\phi),
h_{(F)}^{\mu\nu}(\phi), 0,0,0,0)$.
In construction of suitable generating
functions
$h^{\mu\nu\lambda}=h^{\mu\nu\lambda}(\phi)$, we have to take
into account the dimensions of quantities involved
in the initial action $S_0[\varphi, F, \alpha, \lambda_{(2)}, \lambda_{(4)}]$ (\ref{b1}),
\beq
\nonumber
&&{\rm dim}(\varphi^{\mu\nu\lambda\sigma})=
{\rm dim}(F^{\mu\nu})={\rm dim}(\phi)=\frac{d-2}{2},\\
\label{c3}
&&
{\rm dim}(\xi^{\mu\nu\lambda})=\frac{d-4}{2},\quad {\rm dim}(\pa_{\mu})=1, \;\;
{\rm dim}(\Box)=2.
\eeq
The generating functions $h_{(\varphi)}^{\mu\nu\lambda\sigma}$ and
$h_{(F)}^{\mu\nu}$
should be symmetric and non-local  with
the dimension equals to $-(d+2)/2$.
The non-locality will be achieved by using the operator
$1/\Box$.
To construct cubic vertices $\sim \varphi\phi\phi$,
$h_{(\varphi)}^{\mu\nu\lambda\sigma}$ should be
at least quadratic in fields $\phi$. The tensor structure of
$h_{(\varphi)}^{\mu\nu\lambda\sigma}$
is obeyed by using
partial derivatives $\pa_{\mu}$  and the metric tensor
$\eta_{\mu\nu}$. The minimal number of derivatives equals to 4. Therefore,
the more general form of
$h_{(\varphi)}^{\mu\nu\lambda\sigma}=
h_{(\varphi)}^{\mu\nu\lambda\sigma}(\phi)$ satisfying requirements listed above reads
\beq
\nonumber
&&h_{(\varphi)}^{\mu\nu\lambda\sigma}=a_0\frac{1}{\Box}
\big(c_0\pa^{\mu}\pa^{\nu}\pa^{\lambda}\pa^{\sigma}\phi \;\phi+
c_1\pa^{(\mu}\pa^{\nu}\pa^{\lambda}\phi\;\pa^{\sigma)}\phi+
c_2\pa^{(\mu}\pa^{\nu}\phi\;\pa^{\lambda}\pa^{\sigma)}\phi+\\
\nonumber
&&\qquad\qquad\quad
+c_3\eta^{(\mu\nu}\pa^{\lambda}\pa^{\sigma)}\Box\phi\;\phi+
c_4\eta^{(\mu\nu}\pa^{\lambda}\pa^{\sigma)}\pa_{\rho}\phi\;\pa^{\rho}\phi+
c_5\eta^{(\mu\nu}\pa^{\lambda}\Box\phi\;\pa^{\sigma)}\phi+\\
\label{c4}
&&\qquad\qquad\quad+
c_6\eta^{(\mu\nu}\pa^{\lambda}\pa^{\sigma)}\phi\,\Box\phi+
c_7\eta^{(\mu\nu}\pa_{\rho}\pa^{\lambda}\phi\,\pa^{\sigma)}\pa^{\rho}\phi
\big),
\eeq
where $a_0$ is the coupling constant with ${\rm dim}(a_0)=-d/2$ and
$c_i,\;\; i=0,1,...,7$
are arbitrary dimensionless constants.
By the same reason the generating function $h_{(F)}^{\mu\nu}$ is chosen in the form
\beq
\nonumber
&&h_{(F)}^{\mu\nu}=a_0\frac{1}{\Box}\big(
d_0\Box \pa^{\mu}\pa^{\nu}\phi\;\phi+d_1\pa^{\mu}\pa^{\nu}\pa^{\rho}\phi\;
\pa_{\rho}\phi+d_2\Box\pa^{(\mu}\phi\;\pa^{\nu)}\phi +
d_3\pa^{\mu}\pa^{\nu}\phi\;\Box\phi+\\
\nonumber
&&\qquad\qquad\qquad+d_4\pa_{\rho}\pa^{\mu}\phi\;\pa^{\rho}\pa^{\nu}\phi+
d_5\eta^{\mu\nu}\Box^2\phi\;\phi+
d_6\eta^{\mu\nu}\pa_{\rho}\phi\;\Box\pa^{\rho}\phi+\\
\label{c5}
&&\qquad\qquad\qquad +
d_7\eta^{\mu\nu}\Box \phi\;\Box\phi+
d_8\eta^{\mu\nu}\pa_{\rho}\pa_{\sigma}\phi\;\pa^{\rho}\pa^{\sigma}\phi
\big) ,
\eeq
where $d_i,\;i=0,1,...,8$ are dimensionless constants.

The deformed action has the following explicit and closed form
\beq
\label{c6}
\widetilde{S}[A] = S_0[A]+S_{int}[A],
\eeq
where
\beq
\nonumber
&&S_{int}[A]=2a_0\int dx \Big[\varphi_{\mu\nu\lambda\sigma}K^{\mu\nu\lambda\sigma}-
12F_{\mu\nu}N^{\mu\nu}+\\
\nonumber
&&\qquad\qquad\quad+4\Lambda_{\mu\nu\lambda}\frac{1}{\Box}(\pa_{\sigma}K^{\sigma\mu\nu\lambda}-
2\pa^{(\mu}N^{\nu\lambda)})+
\frac{1}{2}\lambda_{(2)}^{\mu\nu}\frac{1}{\Box}(\eta^{\lambda\sigma}
K_{\lambda\sigma\mu\nu}-
2N_{\mu\nu})+\\
\nonumber
&&\qquad\qquad\quad+\frac{1}{2}\lambda_{(4)}\frac{1}{\Box}\eta^{\mu\nu}N_{\mu\nu}+
\frac{1}{2}a_0K_{\mu\nu\lambda\sigma}\frac{1}{\Box}K^{\mu\nu\lambda\sigma}-
6a_0N_{\mu\nu}\frac{1}{\Box}N^{\mu\nu}+\\
\label{c7}
&&\qquad\qquad\quad+2a_0
\frac{1}{\Box}(\eta^{\lambda\sigma}K_{\lambda\sigma\mu\nu}-2N_{\mu\nu})
\frac{1}{\Box}(\eta_{\rho\beta}
K^{\rho\beta\mu\nu}-
2N^{\mu\nu})\Big]
\eeq
and presentation of generating functions
\beq
\label{c8}
h_{(\varphi)}^{\mu\nu\lambda\sigma}=a_0\frac{1}{\Box}K^{\mu\nu\lambda\sigma},\qquad
h_{(F)}^{\mu\nu}=a_0\frac{1}{\Box}N^{\mu\nu}
\eeq
is used. Notice that $S_{int}[A]$ does not depend on fields $\alpha_{\mu}$.

For Abelian algebras the deformation of gauge symmetry is defined only by the matrix
$M^{-1}(A)$ being inverse to $M(A)=I+h(A)\overleftarrow{\pa}_{\!A}$.
In the sector of fields $(\varphi^{\mu\nu\lambda\sigma}, F^{\mu\nu}, \phi)$
the structure of $M(A)$ looks like
\beq
\label{c9}
\left(\begin{array}{ccc}
E^{\mu\nu\lambda\sigma}_{\rho\beta\gamma\delta}& 0 &
h_{(\varphi)}^{\mu\nu\lambda\sigma}(\phi)\overleftarrow{\pa}_{\!\phi}\\
0&E^{\mu\nu}_{\rho\beta}&h_{(F)}^{\mu\nu}(\phi)\overleftarrow{\pa}_{\!\phi} \\
0&0&1\\
\end{array}\right),
\eeq
In the case of anticanonical transformations described by generating functions
(\ref{c4}) and (\ref{c5}) the inverse matrix $M^{-1}(A)$ can be explicitly found  and
in the sector of fields  $(\varphi^{\mu\nu\lambda\sigma}, F^{\mu\nu}, \phi)$ it
reads
\beq
\label{c10}
\left(\begin{array}{ccc}
E^{\mu\nu\lambda\sigma}_{\rho\beta\gamma\delta}& 0 &
-h_{(\varphi)}^{\mu\nu\lambda\sigma}(\phi)\overleftarrow{\pa}_{\!\phi}\\
0&E^{\mu\nu}_{\rho\beta}&-h_{(F)}^{\mu\nu}(\phi)\overleftarrow{\pa}_{\!\phi} \\
0&0&1\\
\end{array}\right)
.
\eeq
Here,  $E^{\mu\nu\lambda\sigma}_{\rho\beta\gamma\delta}$
and $E^{\mu\nu}_{\rho\beta}$ are elements of the unit matrix in the space of
forth and second  rank symmetric  tensors, respectively. In particular, it means
that in process of deformation of the initial action the gauge transformations
(\ref{c2}) do not deform. In turn, the deformed action
\beq
\label{c11}
\delta\widetilde{S}[A]=0
\eeq
should be  invariant under original gauge transformations (\ref{c2}). As a result,
for the model under consideration we found explicit description of deformed
action and gauge symmetry.

\section{Gauge invariance of local sector of deformed theory}
Now we are in position to study a local sector of the deformed theory. We begin
with the local part of the deformed action
\beq
\label{c12}
S_{loc}[A]=S_0[A]+S_{1\; loc}[\varphi, F, \phi]
\eeq
where $S_{1\; loc}=S_{1\; loc}[\varphi, F, \phi]$,
\beq
\label{c13}
S_{1\; loc}=2a_0\int dx \big[\varphi_{\mu\nu\lambda\sigma}K^{\mu\nu\lambda\sigma}-
12F_{\mu\nu}N^{\mu\nu}\big].
\eeq
Due to the locality of original gauge transformations and
the symmetry property of initial action $S_0[A]$,
from gauge invariance of the deformed action (\ref{c11}) it follows that
the action $S_{1\; loc}[\varphi, F, \phi]$ should be gauge invariant as well,
\beq
\label{c14}
\delta S_{1\; loc}[\varphi, F, \phi]=0.
\eeq
The equation (\ref{c14}) rewrites in the form
\beq
\label{c15}
\int dx \;\xi_{\nu\lambda\sigma}\big[\pa_{\mu}K^{\mu\nu\lambda\sigma}-
3\pa^{\nu}N^{\lambda\sigma}\big] =0,
\eeq
where $\xi_{\nu\lambda\sigma}$ are independent gauge parameters.
Therefore, the equation (\ref{c15}) is equivalent to
\beq
\pa_{\mu}K^{\mu\nu\lambda\sigma}-
\pa^{(\nu}N^{\lambda\sigma)}=0 .
\eeq
Analysis of the
last equation leads to the following relations between constants $c_i, \; i=0,1,...,7$
and $d_j,\; j=0,1,...,8$
\beq
\nonumber
&&c_2=-\frac{1}{4}c_0+c_1,\quad c_5=2c_3,\quad c_7=c_3+c_4-c_6,\\
\nonumber
&&d_0=\frac{1}{3}(c_0+2c_3),\quad
d_1=\frac{1}{3}(c_0+c_1+2c_4),\quad
d_2=-\frac{1}{6}(c_0-3c_1-4c_3),\\
\nonumber
&&d_3=\frac{1}{3}(c_1+2c_6),\quad
d_4=-\frac{1}{3}(c_0-3c_1+2c_3-2c_6),\quad d_5=\frac{2}{3}c_3,\\
\label{c16}
&&d_6=\frac{2}{3}(2c_3+c_4),\quad
d_7=\frac{1}{3}(c_3+c_6),\quad
d_8=\frac{1}{3}(c_3+2c_4-c_6).
\eeq
Therefore, there exists the four-parameter family of local functionals describing cubic
interactions between massless spin 4 and scalar fields and being invariant under original gauge
transformations of the initial free model. These functionals are responsible as well for
local cubic interactions of auxiliary fields $F^{\mu\nu}$ with scalar field $\phi$.

\section{On-shell limit for auxiliary fields}
Next step of our study is to consider
eliminating all auxiliary fields
from the obtained
results using the equations of motion for them. Following \cite{BGK} we extract
$\alpha_{\mu}$ with the help of requirement $\alpha_{\mu}=0$. Then,
from the equations of motion deriving with the help of the action $S_0[A]$ we find
\beq
\label{d1}
F_{\mu\nu}=\frac{1}{2}\eta^{\lambda\sigma}\varphi_{\lambda\sigma\mu\nu},\quad
\eta^{\mu\nu}\eta^{\lambda\sigma}\varphi_{\mu\nu\lambda\sigma}=0,\quad
\eta_{\mu\nu}\xi^{\mu\nu\lambda}=0,\quad
\lambda_{(2)}^{\mu\nu}=0,\quad \lambda_{(4)}=0.
\eeq
In the limit (\ref{d1}) the action (\ref{b1}) reduces to the Fronsdal action for massless
spin 4 fields,
$S_0[\varphi, F, \alpha, \lambda_{(2)}, \lambda_{(4)}]\;\rightarrow \; S_0[\varphi] $,
\beq
\nonumber
&&S_0[\varphi]=\int dx \Big[\varphi_{\mu\nu\lambda\sigma}\Box\varphi^{\mu\nu\lambda\sigma}-
6\eta^{\mu\nu}\eta_{\rho\sigma}
\varphi_{\mu\nu\lambda\delta}\Box\varphi^{\rho\sigma\lambda\delta}-
4\varphi_{\mu\nu\lambda\sigma}\pa^{\mu}\pa_{\rho}\varphi^{\rho\nu\lambda\sigma}+\\
\label{d2}
&&\qquad\qquad\qquad+12\eta^{\mu\nu}\varphi_{\mu\nu\lambda\rho}\pa_{\sigma}\pa_{\delta}
\varphi^{\sigma\delta\lambda\rho}-
6\eta^{\mu\nu}\eta_{\beta\gamma}\varphi_{\mu\nu\lambda\sigma}\pa^{\lambda}\pa_{\rho}
\varphi^{\rho\beta\gamma\sigma}
\Big] .
\eeq
In this limit the local part of the deformed action should be specified by the following
restrictions on parameters of generating functions $K^{\mu\nu\lambda\sigma}$ and $N^{\mu\nu}$
\beq
\nonumber
&&c_2=-\frac{1}{4}c_0+c_1,\quad c_3=c_4=c_5=c_6=c_7=0,\\
&&d_0=\frac{1}{3}c_0,\quad d_1=\frac{1}{3}(c_0+c_1),\quad
d_2=\frac{1}{2}d_4=-\frac{1}{6}(c_0-3c_1),\quad d_3=\frac{1}{3}c_1 .
\eeq
Therefore, the action $S_{1\;loc}[\psi,\phi]$,
\beq
\nonumber
&&S_{1\;loc}[\psi,\phi]=2a_0\int dx \varphi_{\mu\nu\lambda\sigma}\Big[
\pa^{\mu}\pa^{\nu}\pa^{\lambda}\pa^{\sigma}\phi \;\phi+
4c_1\pa^{\mu}\pa^{\nu}\pa^{\lambda}\phi\;\pa^{\sigma}\phi-\\
\nonumber
&&\qquad\qquad\quad-(1-4c_1)\pa^{\mu}\pa^{\nu}\phi\;\pa^{\lambda}\pa^{\sigma}\phi-
2\Box \pa^{\mu}\pa^{\nu}\phi\;\phi\;\eta^{\lambda\sigma}-
2(1+c_1)\pa^{\mu}\pa^{\nu}\pa^{\rho}\phi\; \pa_{\rho}\phi\;\eta^{\lambda\sigma}-\\
&&\qquad\qquad\quad-2c_1\pa^{\mu}\pa^{\nu}\phi\;\Box\phi\; \eta^{\lambda\sigma}+
2(1-3c_1)\pa_{\rho}\pa^{\mu}\phi\;\pa^{\rho}\pa^{\nu}\phi\; \eta^{\lambda\sigma}
\Big],
\eeq
describes the one-parameter family of local actions for massless
constrained spin 4 field
interacting with real scalar field. These actions are   invariant under original
gauge transformations in the Fronsdal theory \cite{Fronsdal-1}.

\section{Discussion}
In the present paper, the new approach \cite{BL-1} was applied
to study interactions between
massless spin 4 ($\varphi^{\mu\nu\lambda\sigma}$) and real scalar ($\phi$) fields.
In contrast with the case of massless spin 3 field
\cite{L-ejpc-s3,L-22-2, L-s5}
it required to use an unconstrained formulation for
free massless integer spin fields of the Fronsdal theory \cite{Fronsdal-1}.
A few years ago such formulation has been proposed \cite{BGK}. Then the initial action
was chosen as the sum of the unconstrained action for massless spin 4 field and the
free action for a real
scalar field. The unconstrained action includes  additionally
in comparison with the Fronsdal action
a set of four auxiliary fields of which two are gauge fields. In the space of all fields
the initial action is gauge invariant with unconstrained gauge parameters
$\xi^{\mu\nu\lambda}$. The gauge algebra is Abelian.

The simplest interactions are cubic ones. In the present study we
were interested in cubic vertices containing interactions of the form
$\sim \varphi\phi\phi$. To construct such kind of vertices using the new method
\cite{BL-1} a more general non-local  anticanonical transformation acting non-trivially
in the space of massless spin 4 field and gauge auxiliary field and containing four partial
derivatives in vertices was proposed (\ref{c4}), (\ref{c5}). The proposed deformation
of initial action leaves the gauge symmetry non-deformed. Then the deformed non-local action
has been described explicitly in the form of functional of fourth order in fields
(\ref{c6}), (\ref{c7}), (\ref{c8}). It was shown that the deformed action contains
a local part (\ref{c12}), (\ref{c13}) which should be invariant under original
gauge transformations (\ref{c14}).
This requirement led to a four-parameter family of cubic vertices containing,
among other things, terms of the form $\sim\varphi\phi\phi$.
In fact, for the first time in the unconstrained approach
to massless integer higher spin fields the family of local gauge invariant actions
with interactions between fields  has been constructed in a closed and explicit form
of fourth-order functionals with respect to  fields.

We analyzed as well on-shell limit for auxiliary fields in
the local deformed unconstrained action constructed in Section 4.
It was found an one-parameter family of cubic vertices invariant under constrained
gauge transformations similarly to gauge invariance of the Fronsdal theory.
Again, we can claim the appearance due to the new method \cite{BL-1}
new examples of consistent local gauge invariant models
with higher spin interactions formulated in terms of constrained fields
of the Fronsdal theory.

\section*{Acknowledgments}
\noindent

The work is supported by the Ministry of
Education of the Russian Federation, project FEWF-2020-0003.

\end{document}